\documentclass[10pt,twocolumn,english,conference]{IEEEtran}
\usepackage[T1]{fontenc}
\usepackage[latin9]{inputenc}
\usepackage{verbatim}
\usepackage{float}
\usepackage{amsmath}
\usepackage{graphicx}
\usepackage{amssymb}

\usepackage{psfrag}

\usepackage{babel}

\begin{document}

\title{Multishot Codes for Network Coding:\\
Bounds and a Multilevel Construction}

\author{\authorblockN{Roberto W. Nóbrega and Bartolomeu F. Uchôa-Filho}
\authorblockA{Communications Research Group\\
Department of Electrical Engineering\\
Federal University of Santa Catarina\\
Florianópolis, SC, 88040-900, Brazil\\
\{rwnobrega, uchoa\}@eel.ufsc.br}}
\maketitle
\begin{abstract}
The subspace channel was introduced by Koetter and Kschischang as
an adequate model for the communication channel from the source node
to a sink node of a multicast network that performs random linear
network coding. So far, attention has been given to one-shot subspace
codes, that is, codes that use the subspace channel only once. In
contrast, this paper explores the idea of using the subspace channel
more than once and investigates the so called multishot subspace codes.
We present definitions for the problem, a motivating example, lower
and upper bounds for the size of codes, and a multilevel construction
of codes based on block-coded modulation.
\end{abstract}

\section{Introduction}

Random linear network coding, first introduced in \cite{ho-benefits},
is an attractive proposal for networks with unknown or changing topology,
in particular for multicast communication, in which there is only
one source but many sink nodes. In this scheme, the network operates
with \emph{packets}, each consisting of $m$~symbols from a finite
field~$\mathbb{F}_{q}$. A packet, then, can be interpreted as a
vector in the vector space~$\mathbb{F}_{q}^{m}$. Each node in the
network transmits random linear combinations of the packets it has
received. As noted in~\cite{koetter-kschischang}, even if the random
coefficients of the linear combinations are not known, it is still
possible to carry out a multicast communication. The key idea is that
the vector subspace spanned by the packets sent by the source node
is preserved over the network and therefore information can be encoded
into subspaces.

Koetter and Kschischang defined in~\cite{koetter-kschischang} the
\emph{subspace channel}, a discrete memoryless channel with input
and output alphabets given by the \emph{projective space}~$\mathcal{P}(\mathbb{F}_{q}^{m})$,\emph{
}which is the collection of all possible vector subspaces of the vector
space~$\mathbb{F}_{q}^{m}$. The source node selects and transmits
an input subspace from the projective space and, in the absence of
errors, the sink nodes receive that same subspace. To deal with the
problem of packet errors and erasures that may happen during the communication,
one can limit the choice of input subspaces to a particular subcollection
of the projective space, i.e., a \emph{subspace code}. Such choice
is driven by a metric known as \emph{subspace distance}, which is
adequate to the subspace channel, according to~\cite{koetter-kschischang}.

We call the codes just described \emph{one-shot subspace codes}, since
they use the subspace channel only once. Many bounds and fundamental
results for one-shot subspace coding, as well as constructions of
codes, have been presented in~\cite{koetter-kschischang,etzion-vardy,gabidulin-bossert}.
In contrast, codes that use the subspace channel many times are called
\emph{multishot subspace codes}, in which the permissible \emph{sequences}
of subspaces to be transmitted are limited to a predetermined subset
of the set of all possible sequences. The present paper explores this
direction.

One of the basic problems in the realm of one-shot subspace coding
is to find codes with good rates and good error correcting/detecting
capabilities. To achieve both goals simultaneously, it may be unavoidable
to increase the field size~$q$ or the packet size~$m$. In view
of that, there are two main reasons that motivate us to consider multishot
subspace coding as an alternative. First, the system under consideration
may be such that it is not possible to change the field and packet
size. And second, even if those parameters are under designer control,
complexity reasons may be determinant---e.g., one-shot codes in $\mathcal{P}(\mathbb{F}_{q}^{mn})$
can be considerably more complicated (although better) than $n$-shot
codes over $\mathcal{P}(\mathbb{F}_{q}^{m})$.

We begin in Section~\ref{sec:definitions} by reviewing definitions
for the one-shot case and introducing new definitions for the multishot
case. In Section~\ref{sec:motivation}, we present a motivation for
multishot coding with a simple example. In Section~\ref{sec:discussion},
we make some pertinent remarks. Section~\ref{sec:relationship} addresses
the relationship between one-shot and multishot codes. Section~\ref{sec:bounds}
derives Hamming-, Gilbert-Varshamov- and Singleton-like bounds for
multishot codes. Section~\ref{sec:multilevel} presents a construction
of multishot codes borrowing ideas from block-coded modulation. Finally,
Section~\ref{sec:conclusion} concludes this paper.

\section{Definitions\label{sec:definitions}}

\subsection{Background}

We start by reviewing some concepts and definitions for one-shot subspace
coding, presented in \cite{koetter-kschischang}.

The \emph{Gaussian binomial} defined by \[
\binom{m}{k}_{q}=\prod_{i=0}^{k-1}\frac{q^{m-i}-1}{q^{k-i}-1}\]
quantifies the the number of $k$-dimensional vector subspaces of
$\mathbb{F}_{q}^{m}$. Therefore, the number of elements in the projective
space~$\mathcal{P}(\mathbb{F}_{q}^{m})$ is given by\[
\left|\mathcal{P}(\mathbb{F}_{q}^{m})\right|=\sum_{k=0}^{m}\binom{m}{k}_{q}.\]

The \emph{subspace distance} between two elements $V$ and $U$ of
the projective space $\mathcal{P}(\mathbb{F}_{q}^{m})$ is defined
as\begin{equation}
d_{\mathrm{S}}(V,U)=\dim(V\dotplus U)-\dim(V\cap U),\label{eq:subspace-distance}\end{equation}
where $V\cap U$ is the intersection of subspaces $V$ and $U$ (which
is clearly a subspace) and $V\dotplus U$ is the sum of subspaces
$V$ and $U$, given by $V\dotplus U=\{v+u:v\in V,\; u\in U\}$ (which
is the smallest subspace containing $V\cup U$). The function $d_{\mathrm{S}}(\cdot,\cdot)$
is indeed a metric over $\mathcal{P}(\mathbb{F}_{q}^{m})$.

In the \emph{subspace channel}, we transmit a subspace $V\in\mathcal{P}(\mathbb{F}_{q}^{m})$
and receive another subspace $U\in\mathcal{P}(\mathbb{F}_{q}^{m})$.
If $V\neq U$, an error has occurred. The \emph{weight} of the error
is defined as $d_{\mathrm{S}}(V,U)$. We call an error of weight~$1$
a \emph{single error}, an error of weight~$2$ a \emph{double error},
and so on.

\subsection{Multishot Subspace Coding}

We now introduce definitions for the multishot case by considering
\emph{block codes} of lenght~$n$ over a projective space. In other
words, we consider codes in which the subspace channel just defined
is used $n$~times.

The $n$th \emph{extension of the projective space} $\mathcal{P}(\mathbb{F}_{q}^{m})$
is defined as $\mathcal{P}(\mathbb{F}_{q}^{m})^{n}$, that is, the
$n$th Cartesian power of the projective space. Thus, elements of
$\mathcal{P}(\mathbb{F}_{q}^{m})^{n}$ are $n$-tuples of subspaces
in $\mathcal{P}(\mathbb{F}_{q}^{m})$. Of course, the number of elements
in $\mathcal{P}(\mathbb{F}_{q}^{m})^{n}$ is given by\[
\left|\mathcal{P}(\mathbb{F}_{q}^{m})^{n}\right|=\left|\mathcal{P}(\mathbb{F}_{q}^{m})\right|^{n}.\]

The \emph{extended subspace distance} between two elements $\mathbf{V}=(V_{1},\ldots,V_{n})$
and $\mathbf{U}=(U_{1},\ldots,U_{n})$ of $\mathcal{P}(\mathbb{F}_{q}^{m})^{n}$
is defined as\begin{equation}
d_{\mathrm{S}}(\mathbf{V},\mathbf{U})=\sum_{i=0}^{n}d_{\mathrm{S}}(V_{i},U_{i}),\label{eq:subspace-distance-ms}\end{equation}
where $d_{\mathrm{S}}(\cdot,\cdot)$ in the right-hand side is given
by~\eqref{eq:subspace-distance}.

Here, we transmit a $n$-tuple of subspaces $\mathbf{V}=(V_{1},\ldots,V_{n})$
and receive another $n$-tuple of subspaces $\mathbf{U}=(U_{1},\ldots,U_{n})$.
In the absence of errors, $\mathbf{V}=\mathbf{U}$. Otherwise, an
error of \emph{total weight} $d_{\mathrm{S}}(\mathbf{V},\mathbf{U})$
has occurred. We note that, for example, two single errors occurring
in different transmissions amounts to one double error occurring in
some transmission, since both cases gives a total weight of~$2$.

A \emph{multishot (block) subspace code of length}~$n$ (also called
a \emph{$n$-shot subspace code}) over $\mathcal{P}(\mathbb{F}_{q}^{m})$
is a non-empty subset of $\mathcal{P}(\mathbb{F}_{q}^{m})^{n}$. The
\emph{size} of a code~$\mathcal{C}$ is given by~$\left|\mathcal{C}\right|$,
and the \emph{rate} of that code is defined as

\[
R(\mathcal{C})=\frac{\log\left|\mathcal{C}\right|}{n},\]
measured in information symbols per subspace channel use. Finally,
the \emph{minimum distance} of~$\mathcal{C}$ is defined as\[
d_{\mathrm{S}}(\mathcal{C})=\min\{d_{\mathrm{S}}(\mathbf{V},\mathbf{U}):\mathbf{U},\mathbf{V}\in\mathcal{C},\;\mathbf{U}\neq\mathbf{V}\}.\]

We have $1\leq d_{\mathrm{S}}(\mathcal{C})\leq mn$ and $0\leq R(\mathcal{C})\leq1$,
if the logarithm base is taken as $\left|\mathcal{P}(\mathbb{F}_{q}^{m})\right|$.

\section{A Motivating Example\label{sec:motivation}}

\begin{figure}[H]
\begin{centering}
\hspace{-1cm}\psfrag{O}[cc][cc][1.0][0]{$O$}
\psfrag{W}[cc][cc][1.0][0]{$W$}
\psfrag{V1}[cc][cc][1.0][0]{$S_1$}
\psfrag{V2}[cc][cc][1.0][0]{$S_2$}
\psfrag{V3}[cc][cc][1.0][0]{$S_3$}
\psfrag{LEGEND1}[cc][cc][1.0][0]{$O=\{00\}$}
\psfrag{LEGEND2}[cc][cc][1.0][0]{$S_1=\{00,01\}$}
\psfrag{LEGEND3}[cc][cc][1.0][0]{$S_2=\{00,10\}$}
\psfrag{LEGEND4}[cc][cc][1.0][0]{$S_3=\{00,11\}$}
\psfrag{LEGEND5}[cc][cc][1.0][0]{$W=\{00,01,10,11\}$}\includegraphics[scale=0.67]{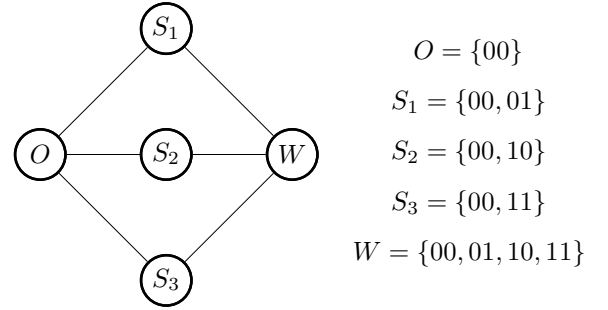}\caption{Projective space $\mathcal{P}(\mathbb{F}_{2}^{2})$.}

\par\end{centering}

\centering{}\label{fig:hasse-F22}
\end{figure}

Suppose we wish a multishot subspace code using the projective space
$\mathcal{P}(\mathbb{F}_{2}^{2})$ whose Hasse graph~\cite{koetter-kschischang}
is shown in Figure~\ref{fig:hasse-F22}. Suppose also that our goal
is to be able to detect a single error occurring in any of the $n=3$
transmissions.%
\footnote{That is, we are considering an adversarial error model in which at~most
a single error can occur in a block of $3$~transmissions.%
} So, it suffices to find a $3$-shot code with minimum distance $d=2$.

A first approach is simply to extend the best one-shot subspace code
in $\mathcal{P}(\mathbb{F}_{2}^{2})$ with minimum distance~$2$,
which is \[
\mathcal{C}_{1}'=\{S_{1},S_{2},S_{3}\}.\]
By doing so we obtain the code\begin{eqnarray*}
\mathcal{C}_{1} & = & \mathcal{C}_{1}'\times\mathcal{C}_{1}'\times\mathcal{C}_{1}'\\
 & = & \{S_{1}S_{1}S_{1},S_{1}S_{1}S_{2},S_{1}S_{1}S_{3},\ldots,S_{3}S_{3}S_{3}\}\end{eqnarray*}
with $\left|\mathcal{C}_{1}\right|=27$.

Can we do better? Let us try to consider the projective space $\mathcal{P}(\mathbb{F}_{2}^{2})$
as an alphabet of a \textquotedblleft{}classical\textquotedblright{}
code. Accordingly, take any bijective mapping between $\mathcal{P}(\mathbb{F}_{2}^{2})=\{O,S_{1},S_{2},S_{3},W\}$
and $\mathbb{Z}_{5}=\{0,1,2,3,4\}$, for example, $O\mapsto0$, $S_{1}\mapsto1$,
$S_{2}\mapsto2$, $S_{3}\mapsto3$ and $W\mapsto4$. The best classical
code of length~$3$ over~$\mathbb{Z}_{5}$ with minimum Hamming
distance~$2$ is a parity-check code, such as \begin{eqnarray*}
C_{2} & = & \{x_{1}x_{2}x_{3}\in\mathbb{Z}_{5}^{3}:x_{1}+x_{2}+x_{3}=0\}\\
 & = & \{000,014,023,\ldots,442\},\end{eqnarray*}
which is mapped back to \[
\mathcal{C}_{2}=\{OOO,OS_{1}W,OS_{2}S_{3},\ldots,WWS_{2}\}\]
with $\left|\mathcal{C}_{2}\right|=25$, smaller than~$\left|\mathcal{C}_{1}\right|$.

The second approach did not succeed because it disregarded the subspace
structure behind $\mathcal{P}(\mathbb{F}_{2}^{2})$ and used only
classical coding. If we want to achieve better results, we must, in
fact, design codes in the metric space~$\mathcal{P}(\mathbb{F}_{2}^{2})^{3}$,
taking into account both the subspace structure and time evolution.
In Section~\ref{sec:multilevel}, following this idea, we find a
code~$\mathcal{C}_{3}$ in $\mathcal{P}(\mathbb{F}_{2}^{2})^{3}$
with minimum distance~$2$ and $\left|\mathcal{C}_{3}\right|=63$
by means of a multilevel construction.

\section{Some Remarks on Multishot Codes\label{sec:discussion}}

\subsection{Rate of a Code}

In Section~\ref{sec:definitions}, we have defined the rate of a
code~$\mathcal{C}$ as $R(\mathcal{C})=\frac{1}{n}\log\left|\mathcal{C}\right|$,
measured in information symbols per subspace channel use. However,
such definition may not be suitable for all situations. A good definition
for rate is one which captures the notion of {}``cost'' for the
transmission of codewords. Although information is coded into subspaces,
in practice we transmit vectors (packets) that form a basis for the
subspace and not the subspace itself.

With this in mind, and following the work in~\cite{koetter-kschischang},
it may be interesting to redefine the rate of~$\mathcal{C}$ either
as $R(\mathcal{C})=\frac{1}{\ell(\mathcal{C})\cdot n}\log\left|\mathcal{C}\right|$,
measured in information symbols per packet transmitted, or $R(\mathcal{C})=\frac{1}{m\cdot\ell(\mathcal{C})\cdot n}\log\left|\mathcal{C}\right|$,
measured in information symbols per $q$-ary symbol transmitted. In
the definitions, the quantity $\ell(\mathcal{C})$ can be either the
average or the maximum dimension of the subspaces in code~$\mathcal{C}$.
This is specially valid for a generation-based model~\cite{chou},
in which \textquotedblleft{}to transmit a subspace would require the
transmitter to inject on average (or up to) $\ell(\mathcal{C})$ packets
into the network, corresponding to the transmission of $m\!\cdot\!\ell(\mathcal{C})$
$q$-ary symbols\textquotedblright{}, still according to~\cite{koetter-kschischang}.

\subsection{Error Control Capability of a Code}

Similarly to classical codes, multishot subspace codes with minimum
distance~$d$ can detect every error of total weight~$d-1$ or less
and correct every error of total weight~$\left\lfloor (d-1)/2\right\rfloor $
or less. So, is code~$\mathcal{C}_{3}$ of Section~\ref{sec:motivation}
better than code~$\mathcal{C}_{1}$? If all we require is to detect
a single error in any of the $3$~transmissions, the answer is affirmative,
since both can certainly detect a single error and code~$\mathcal{C}_{3}$
has a larger number of codewords. But code~$\mathcal{C}_{1}$ can
detect $3$~errors, as long as each of them occur in a different
transmission.%
\footnote{Even so, we cannot call code~$\mathcal{C}_{1}$ a \textquotedblleft{}$3$-error-detecting
code\textquotedblright{}, since it cannot detect \emph{all} errors
of total weight~$3$ or less (e.g., it cannot detect a double error
occurring in any transmission).%
} In view of that, the \emph{normalized distance}~$d_{\mathrm{S}}(\mathcal{C})/n$
may be a better parameter to settle when comparing two multishot codes.
For example, code~$\mathcal{C}_{1}'$, the one-shot counterpart of
code~$\mathcal{C}_{1}$, has normalized distance~$2$, while code~$\mathcal{C}_{3}$
has normalized distance~$2/3$.

The purpose of the foregoing discussion was to emphasize the significance
of the error model being adopted. Besides that, another important
subject is the relation of subspace errors to packet errors and erasures.
Such study is made in~\cite{koetter-kschischang,silva-metrics} for
one-shot subspace coding and could be extended to the multishot case.

\section{Relationship to One-Shot Codes\label{sec:relationship}}

Obviously, one-shot codes are just a special case of $n$-shot codes---just
set $n=1$. In this section, we show how the converse statement can
also be interpreted to be true in a sense.

The $n$th extension of a projective space, $\mathcal{P}(\mathbb{F}_{q}^{m})^{n}$,
can be viewed as a \textquotedblleft{}subset\textquotedblright{} of
the larger projective space $\mathcal{P}(\mathbb{F}_{q}^{mn})$. To
see how, consider an injective mapping $f:\mathcal{P}(\mathbb{F}_{q}^{m})^{n}\longrightarrow\mathcal{P}(\mathbb{F}_{q}^{mn})$
defined as follows. Let $\mathbf{V}=(V_{1},\ldots,V_{n})\in\mathcal{P}(\mathbb{F}_{q}^{m})^{n}$
and let $\mathbf{b}_{i,1},\ldots,\mathbf{b}_{i,m}\in\mathbb{F}_{q}^{m}$
be vectors such that $V_{i}=\left\langle \mathbf{b}_{i,1},\ldots,\mathbf{b}_{i,m}\right\rangle $
(i.e., the vector space spanned by $\mathbf{b}_{i,1},\ldots,\mathbf{b}_{i,m}$),
for $i=1,\ldots,n$. Then, $f$ is defined as \begin{eqnarray*}
f(\mathbf{V}) & = & \langle(\mathbf{b}_{1,1},\mathbf{0},\ldots,\mathbf{0}),\ldots,(\mathbf{b}_{1,m},\mathbf{0},\ldots,\mathbf{0}),\\
 &  & \ (\mathbf{0},\mathbf{b}_{2,1},\ldots,\mathbf{0}),\ldots,(\mathbf{0},\mathbf{b}_{2,m},\ldots,\mathbf{0}),\\
 &  & \qquad\qquad\qquad\qquad\vdots\\
 &  & \ (\mathbf{0},\ldots,\mathbf{0},\mathbf{b}_{n,1}),\ldots,(\mathbf{0},\ldots,\mathbf{0},\mathbf{b}_{n,m})\rangle.\end{eqnarray*}

It can be shown that $f$ is really injective and that $d_{\mathrm{S}}(\mathbf{V},\mathbf{U})=d_{\mathrm{S}}(f(\mathbf{V}),f(\mathbf{U}))$
for every $\mathbf{V},\mathbf{U}\in\mathcal{P}(\mathbb{F}_{q}^{m})^{n}$.
So, every $n$-shot code~$\mathcal{C}\subseteq\mathcal{P}(\mathbb{F}_{q}^{m})^{n}$
leads to an one-shot code~$f(\mathcal{C})\subseteq\mathcal{P}(\mathbb{F}_{q}^{mn})$
with same minimum distance and size.

This also suggests a construction for multishot codes in $\mathcal{P}(\mathbb{F}_{q}^{m})^{n}$
based on one-shot codes in $\mathcal{P}(\mathbb{F}_{q}^{mn})$. Indeed,
if we take a code $\mathcal{C}\subseteq\mathcal{P}(\mathbb{F}_{q}^{mn})$
with minimum distance~$d$ and throw away the codewords that are
not in $f(\mathcal{P}(\mathbb{F}_{q}^{m})^{n})$, we get a code~$\mathcal{C}'$,
and $f^{-1}(\mathcal{C}')\subseteq\mathcal{P}(\mathbb{F}_{q}^{m})^{n}$
is a $n$-shot code with minimum distance at least $d$, but with
a lower rate. Yet, it is not clear if good codes in $\mathcal{P}(\mathbb{F}_{q}^{mn})$
always lead to good codes in $\mathcal{P}(\mathbb{F}_{q}^{m})^{n}$.

\section{Bounds on Codes\label{sec:bounds}}

Let $\mathcal{A}_{q}^{n}(m,d)$ denote the size of the largest code
in~$\mathcal{P}(\mathbb{F}_{q}^{m})^{n}$ with minimum distance~$d$,
that is, \[
\mathcal{A}_{q}^{n}(m,d)=\max\{\left|\mathcal{C}\right|:\mathcal{C}\subseteq\mathcal{P}(\mathbb{F}_{q}^{m})^{n}\mbox{ and }d_{\mathrm{S}}(\mathcal{C})=d\}.\]
In this section we derive upper and lower bounds on $\mathcal{A}_{q}^{n}(m,d)$.

Of course, every lower bound for $\left|\mathcal{P}(\mathbb{F}_{q}^{m})\right|$-ary
classical codes is a lower bound on $\mathcal{A}_{q}^{n}(m,d)$, a
fact following from the discussion in Section~\ref{sec:motivation}.
Likewise, every upper bound for one-shot codes in $\mathcal{P}(\mathbb{F}_{q}^{mn})$
is an upper bound on $\mathcal{A}_{q}^{n}(m,d)$, according to Section~\ref{sec:relationship}.
Hence, \[
A_{\left|\mathcal{P}(\mathbb{F}_{q}^{m})\right|}(n,d)\leq\mathcal{A}_{q}^{n}(m,d)\leq\mathcal{A}_{q}(mn,d),\]
where $A_{q'}(n,d)$ is the size of the best classical code of length~$n$
over $\mathbb{F}_{q'}$ with minimum Hamming distance~$d$ and $\mathcal{A}_{q}(m',d)=\mathcal{A}_{q}^{1}(m',d)$
is the size of the best one-shot code in $\mathcal{P}(\mathbb{F}_{q}^{m'})$
with minimum subspace distance~$d$.

\subsection{Sphere-Packing and Sphere-Covering Bounds}

For the next two bounds we will need the notion of spheres lying in
the metric space~$\mathcal{P}(\mathbb{F}_{q}^{m})^{n}$. The \emph{sphere}
centered in $\mathbf{V}=(V_{1},\ldots,V_{n})$ with radius~$r$ in
$\mathcal{P}(\mathbb{F}_{q}^{m})^{n}$ is given by\[
\mathcal{B}_{(q,m,n)}(\mathbf{V},r)=\{\mathbf{U}\in\mathcal{P}(\mathbb{F}_{q}^{m})^{n}:d_{\mathrm{S}}(\mathbf{U},\mathbf{V})\leq r\},\]
and the \emph{volume} of that sphere is defined as\[
\mathrm{Vol}_{(q,m,n)}(\mathbf{V},r)=\left|\mathcal{B}_{(q,m,n)}(\mathbf{V},r)\right|.\]

It can be shown that\[
\mathrm{Vol}_{(q,m,n)}(\mathbf{V},r)=\sum_{\substack{\mathbf{j}\in\{0,\ldots,m\}^{n}:\\
j_{1}+\cdots+j_{n}\leq r}
}\;\;\prod_{i=1}^{n}\mathrm{Vol}_{(q,m)}^{\mathrm{Shell}}(V_{i},j_{i}),\]
where

\pagebreak{}\[
\mathrm{Vol}_{(q,m)}^{\mathrm{Shell}}(V,j)=\sum_{i=0}^{j}\binom{m-k}{j-i}_{q}\binom{k}{i}_{q}q^{i(j-i)}\]
is the volume of a shell of subspaces with radius $j$ centered in
$V$ with $\dim V=k$ in the projective space $\mathcal{P}(\mathbb{F}_{q}^{m})$,
as given in \cite{koetter-kschischang,etzion-vardy}.

The volume of a shell centered in~$\mathbf{V}$ depends only on $\mathbf{k}=(\dim V_{1},\ldots,\dim V_{n})$,
so we also adopt the notation $\mathrm{Vol}_{(q,m,n)}(\mathbf{k},r)$.
Moreover, we will drop the subscripts for convenience.

%
{}

Given a tuple $\mathbf{k}=(k_{1},\ldots,k_{n})$, there are a total
of\[
\mathrm{Freq}_{(q,m,n)}(\mathbf{k})=\binom{m}{k_{1}}_{q}\cdots\binom{m}{k_{n}}_{q}\]
points $\mathbf{V}$ such that $\mathbf{k}=(\dim V_{1},\ldots,\dim V_{n})$.
Therefore, the average volume of a sphere of radius $r$ in $\mathcal{P}(\mathbb{F}_{q}^{m})^{n}$
is\begin{eqnarray}
\mathrm{Vol}^{\mathrm{avg}}(r) & = & \frac{1}{\left|\mathcal{P}(\mathbb{F}_{q}^{m})^{n}\right|}\sum_{\mathbf{V}\in\mathcal{P}(\mathbb{F}_{q}^{m})^{n}}\mathrm{Vol}(\mathbf{V},r)\label{eq:vol-avg}\\
 & = & \frac{1}{\left|\mathcal{P}(\mathbb{F}_{q}^{m})\right|^{n}}\sum_{\mathbf{k}\in\{1,\ldots,m\}^{n}}\mathrm{Freq}(\mathbf{k})\mathrm{Vol}(\mathbf{k},r).\nonumber \end{eqnarray}
Also, the maximum and minimum volumes are\begin{eqnarray}
\mathrm{Vol}^{\mathrm{min}}(r) & = & \mathrm{Vol}((\left\lfloor m/2\right\rfloor ,\ldots,\left\lfloor m/2\right\rfloor ),r),\label{eq:vol-min}\\
\mathrm{Vol}^{\mathrm{max}}(r) & = & \mathrm{Vol}((0,\ldots,0),r).\label{eq:vol-max}\end{eqnarray}

If we consider the packing of spheres of radius~$r=\left\lfloor (d-1)/2\right\rfloor $
centered at the codewords of a code~$\mathcal{C}$ in $\mathcal{P}(\mathbb{F}_{q}^{m})^{n}$,
we get\begin{eqnarray*}
\left|\mathcal{P}(\mathbb{F}_{q}^{m})\right|^{n} & \geq & \sum_{\mathbf{V}\in\mathcal{C}}\mathrm{Vol}(\mathbf{V},r)\\
 & \geq & \sum_{\mathbf{V}\in\mathcal{C}}\mathrm{Vol}^{\mathrm{min}}(r)\\
 & = & \left|\mathcal{C}\right|\mathrm{Vol}^{\mathrm{min}}(r),\end{eqnarray*}
and so we have the Hamming-like upper bound\[
\mathcal{A}_{q}^{n}(m,d)\leq\frac{\left|\mathcal{P}(\mathbb{F}_{q}^{m})\right|^{n}}{\mathrm{Vol}^{\mathrm{min}}(\left\lfloor (d-1)/2\right\rfloor )},\]
where $\mathrm{Vol}^{\mathrm{min}}(\cdot)$ is given by \eqref{eq:vol-min}.

The same approach used in \cite{etzion-vardy} for the one-shot case
can be used here to get the Gilbert-Varshamov-like lower bound\[
\mathcal{A}_{q}^{n}(m,d)\geq\frac{\left|\mathcal{P}(\mathbb{F}_{q}^{m})\right|^{n}}{\mathrm{Vol}^{\mathrm{avg}}(d-1)},\]
where $\mathrm{Vol}^{\mathrm{avg}}(\cdot)$ is given by \eqref{eq:vol-avg}.

\subsection{Singleton Bound}

We now consider a puncturing operation of a codeword \begin{eqnarray*}
(\cdot)^{\blacktriangledown}:\mathcal{P}(\mathbb{F}_{q}^{m})^{n} & \longrightarrow & \mathcal{P}(\mathbb{F}_{q}^{m})^{n-1}\\
\mathbf{V} & \longmapsto & \mathbf{V}^{\blacktriangledown},\end{eqnarray*}
which consists in removing any coordinate of tuple $\mathbf{V}$.
The punctured code is defined as $\mathcal{C}^{\blacktriangledown}=\{\mathbf{V}^{\blacktriangledown}:\mathbf{V}\in\mathcal{C}\}$.
One can prove that if $d_{\mathrm{S}}(\mathcal{C})>m$ then $|\mathcal{C}^{\blacktriangledown}|=|\mathcal{C}|$
and $d_{\mathrm{S}}(\mathcal{C}^{\blacktriangledown})\geq d_{\mathrm{S}}(\mathcal{C})-m$.

Let $\mathcal{C}\subseteq\mathcal{P}(\mathbb{F}_{q}^{m})^{n}$ be
a code with $d_{\mathrm{S}}(\mathcal{C})=d$. By puncturing the code
$\left\lfloor \frac{d-1}{m}\right\rfloor $ times we get a code $\mathcal{C}'=\mathcal{C}^{\blacktriangledown\cdots\blacktriangledown}\subseteq\mathcal{P}(\mathbb{F}_{q}^{m})^{n-\left\lfloor \frac{d-1}{m}\right\rfloor }$
with $|\mathcal{C}^{'}|=|\mathcal{C}|$ and $d_{\mathrm{S}}(\mathcal{C}^{'})\geq1$.
Therefore the Singleton-like upper bound becomes\[
\mathcal{A}_{q}^{n}(m,d)\leq\left|\mathcal{P}(\mathbb{F}_{q}^{m})\right|^{n-\left\lfloor \frac{d-1}{m}\right\rfloor }.\]

\section{Multilevel Construction\label{sec:multilevel}}

In this section, we propose a method for constructing multishot codes
which is inspired by the so-called multilevel construction for block-coded
modulation schemes \cite{imai-hirakawa,calderbank}. This code construction
was first proposed by Imai and Hirakawa~\cite{imai-hirakawa} in~1977,
and became very popular in the 80's and 90's with more general constructions
being developed by many other researchers. Next, we base our description
of the multilevel construction on the work of Calderbank \cite{calderbank},
wherein many references on this subject are listed.

Given an initial set~$\Gamma_{0}$, an \emph{$L$-level partition}
is defined as a sequence of partitions $\Gamma_{0},\ldots,\Gamma_{L}$,
where the partition~$\Gamma_{l}$ is a refinement of $\Gamma_{l-1}$,
in the sense that the subsets in $\Gamma_{l}$ are subsubsets of the
subsets in $\Gamma_{l-1}$. The simplest way to perform an $L$-level
partition is to construct a rooted tree with $L+1$~levels where
the root is the initial set~$\Gamma_{0}$ and the vertices at level~$l$
are the subsets in the partition~$\Gamma_{l}$. In the tree, a subset~$\mathcal{Y}$
in $\Gamma_{l}$ at level~$l$ is joined to the unique subset~$\mathcal{X}$
in $\Gamma_{l-1}$ at level~$l-1$ containing $\mathcal{Y}$, and
to every subset~$\mathcal{Z}$ in $\Gamma_{l+1}$ at level~$l+1$
that is contained in $\mathcal{Y}$. The leaves (i.e., the elements
of $\Gamma_{L}$ at level~$L$) correspond to all the elements of
$\Gamma_{0}$ viewed individually as subsets.

In our construction of multishot subspace codes, we must require \emph{nested
partitions} up to a certain level of the tree. A partition, say $\Gamma_{l}$
at level $l\geq1$, is a nested partition if every subset in $\Gamma_{l-1}$
is joined to the same number $p_{l}$ of subsets in $\Gamma_{l}$,
although we do allow the subsets in $\Gamma_{l}$ to have different
cardinalities. The edges used to join a subset at level $l-1$ to
subsets at level $l$ in the tree can then be labeled with the numbers
$0,\ldots,p_{l}-1$. With this labeling, the subsets in $\Gamma_{l}$
at level $l$ can be labeled by paths $(a_{1},\ldots,a_{l})$, where
$a_{i}\in\{0,\ldots,p_{i}-1\}$.

We start our construction by forming an $L$-level partition of the
entire projective space $\Gamma_{0}=\mathcal{P}(\mathbb{F}_{q}^{m})$.
The metric in this case is the subspace distance defined in~\eqref{eq:subspace-distance}.
We define the \emph{intrasubset (subspace) distance} $d_{\mathrm{S}}^{(l)}$
of level $l$ as\[
d_{\mathrm{S}}^{(l)}=\min_{\mathcal{S}\in\Gamma_{l}}\{d_{\mathrm{S}}(U,V):U,V\in\mathcal{S},\ U\neq V\},\]
for $l=0,\ldots,L$.

\begin{figure}[H]
\centering{}\psfrag{0}[cc][cc][0.8][0]{$0$}
\psfrag{1}[cc][cc][0.8][0]{$1$}
\psfrag{2}[cc][cc][0.8][0]{$2$}\includegraphics[scale=0.33]{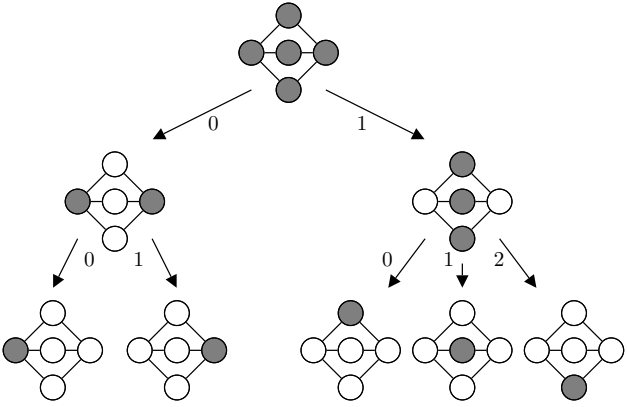}\caption{Example for multilevel construction.}
\label{fig:multilevel-F22}
\end{figure}

Figure~\ref{fig:multilevel-F22} shows an example of a $2$-level
partitioning starting with $\Gamma_{0}=\mathcal{P}(\mathbb{F}_{2}^{2})$.
We have $d_{\mathrm{S}}^{(0)}=1$, $d_{\mathrm{S}}^{(1)}=2$, $d_{\mathrm{S}}^{(2)}=\infty$.
It should be noticed that partition~$\Gamma_{1}$ is nested, while
partition~$\Gamma_{2}$ is not.

We want to construct a $n$-shot subspace code $\mathcal{C}\subseteq\mathcal{P}(\mathbb{F}_{q}^{m})^{n}$
with minimum distance $d_{\mathrm{S}}(\mathcal{C})=d$. We first form
a multilevel partition of $\Gamma_{0}=\mathcal{P}(\mathbb{F}_{q}^{m})$,
and then find the corresponding intrasubset distances. Say we find
that $L^{\prime}$ is the minimum level satisfying $d_{\mathrm{S}}^{(l)}\geq d$
for all~$l\geq L'$. We have to make sure that all partitions up
to level $L^{\prime}$ are nested partitions, throwing out subspaces
if necessary. Then, we must find classical block codes (called \emph{component
codes}) $C_{l}\subseteq\mathbb{Z}_{p_{l}}^{n},$ $1\leq l\leq L^{\prime}$,
with maximal rates and minimum Hamming distance $d_{\mathrm{H}}^{(l)}$
such that $\min\{d_{\mathrm{S}}^{(l-1)}d_{\mathrm{H}}^{(l)}:1\leq l\leq L^{\prime}\}\geq d$.

The codewords of the $n$-shot subspace code $\mathcal{C}\subseteq\mathcal{P}(\mathbb{F}_{q}^{m})^{n}$
are obtained as follows. Form $\prod_{l=1}^{L^{\prime}}|C_{l}|$ arrays
of $L^{\prime}$ rows and $n$ columns, where each array is formed
by arranging a codeword of code $C_{l}$ in its $l$-th row. Let $\mathcal{A}$
denote the collection of all such arrays. Let the $i$-th column of
array~$\mathbf{A}\in\mathcal{A}$ be denoted by $(a_{1,i},\ldots,a_{L^{\prime},i})^{T}$.
A length-$n$ codeword of the subspace code $\mathcal{C}$ is formed
by selecting for its $i$-th coordinate a subspace from the subset
of $\Gamma_{L^{\prime}}$ at level $L^{\prime}$ whose label is the
path $(a_{1,i},\ldots,a_{L^{\prime},i})$. If there are $|\Gamma_{L^{\prime}}(a_{1,i},\ldots,a_{L^{\prime},i})|$
such subspaces, then the number of codewords in the $n$-shot subspace
code $\mathcal{C}\subseteq\mathcal{P}(\mathbb{F}_{q}^{m})^{n}$ is
given by\begin{equation}
|\mathcal{C}|=\sum_{\mathbf{A}\in\mathcal{A}}\left(\prod_{i=1}^{n}|\Gamma_{L^{\prime}}(a_{1,i},\ldots,a_{L^{\prime},i})|\right).\label{eq:multilevel-cardinality}\end{equation}
Also, it is guaranteed that the minimum distance of $\mathcal{C}$
is \begin{equation}
d_{\mathrm{S}}(\mathcal{C})\geq\min\{d_{\mathrm{S}}^{(l-1)}d_{\mathrm{H}}^{(l)}:1\leq l\leq L^{\prime}\}.\label{eq:multilevel-distance}\end{equation}

Back to our example of Figure~\ref{fig:multilevel-F22}, suppose
we wish to construct a $3$-shot subspace code with minimum distance~$2$,
which implies $L'=1$. From \eqref{eq:multilevel-distance}, we must
find a binary ($p_{1}=2$) classical code~$C_{1}$ with $d_{\mathrm{H}}(C_{1})=d_{\mathrm{H}}^{(1)}\geq2$.
The best binary classical codes with lenght~$3$ and minimum distance~$2$
are the even parity-bit code~$C_{1}=\{000,011,101,110\}$ and its
coset, the odd parity-bit code~$C_{1}'=\{001,010,100,111\}$. The
multilevel construction using using $C_{1}$ (resp., $C_{1}'$) gives
a $3$-shot subspace code with minimum distance $2$ and $62$ (resp.,
$63$) codewords.

\section{Conclusion\label{sec:conclusion}}

The aim of this paper was to suggest multishot subspace coding as
a potential alternative to one-shot subspace coding, specially when
the field size~$q$ or packet size~$m$ cannot be changed. Multishot
subspace coding introduces a new degree of freedom: the number of
channel uses~$n$.

Future directions of research may include the following.
\begin{enumerate}
\item The use of \emph{convolutional coding} instead of block coding by
considering ideas similar to Ungerboeck's \emph{trellis-coded modulation}
\cite{ungerboeck}.
\item The determination of the \emph{subspace channel capacity} under a
probabilistic error model and an information-theoretical point of
view. The works \cite{montanari-urbanke,silva-capacity} deal with
the so called \textquotedblleft{}one-shot capacity\textquotedblright{}
and find assymptotical expressions when either the symbol size or
packet size (or both) increases.
\item Finally, the development of bounds and constructions for \emph{constant-dimension}%
\footnote{Constant-dimension subspace codes are codes that contain only subspaces
of a given dimension. Those are also called {}``codes in the Grassmannian'',
since the collection of all vector subspaces with a given dimension
is called a \emph{Grassmannian}.%
}\emph{ multishot subspace codes}. For the one-shot case, refer to~\cite{koetter-kschischang,etzion-vardy,gabidulin-bossert}
and~\cite{silva-rank-metric,etzion-silberstein}, the last two based
on a related metric called the \emph{rank-metric}.
\end{enumerate}

\section*{Acknowledgment}

The authors would like to thank CAPES (Brazil) and CNPq (Brazil) for
the financial support.

\bibliographystyle{ieeetr}
\bibliography{biblio}

\end{document}